\documentclass[
    ,final              ]{aipproc}

\layoutstyle{8x11single}
\begin{document}

\title{Do supernovae favor tachyonic Big Brake instead de Sitter ?}


\author{L. \'{A}. Gergely}{
  address={Department of Theoretical Physics, University of Szeged, Tisza Lajos krt
84-86, Szeged 6720, Hungary },altaddress={
Department of Experimental Physics, University of Szeged, D\'{o}m 
T\'{e}r 9, Szeged 6720, Hungary}
}

\author{Z. Keresztes}{
  address={Department of Theoretical Physics, University of Szeged, Tisza Lajos krt
84-86, Szeged 6720, Hungary },altaddress={
Department of Experimental Physics, University of Szeged, D\'{o}m 
T\'{e}r 9, Szeged 6720, Hungary}
}

\author{ A. Yu. Kamenshchik}{
  address={Dipartimento di Fisica and INFN, via Irnerio 46, 40126 Bologna, Italy },altaddress={
L.D. Landau Institute for Theoretical Physics, Russian Academy of Sciences,
Kosygin street 2, 119334 Moscow, Russia}
}

\author{V. Gorini}{
  address={Dipartimento di Scienze Fisiche e Mathematiche, Universit\`a dell'Insubria, Via Valleggio 11, 22100 Como, Italy },altaddress={
INFN, sez. di Milano, Via Celoria 16, 20133 Milano, Italy}
}

\author{U. Moschella}{
  address={Dipartimento di Scienze Fisiche e Mathematiche, Universit\`a dell'Insubria, Via Valleggio 11, 22100 Como, Italy },altaddress={
INFN, sez. di Milano, Via Celoria 16, 20133 Milano, Italy}
}

\begin{abstract}
We investigate whether a tachyonic scalar field, encompassing both dark energy and dark matter-like features will drive our universe towards a Big Brake singularity or a de Sitter expansion. In doing this it is crucial to establish the parameter domain of the model, which is compatible with type Ia supernovae data. We find the 1$\sigma$ contours and evolve the tachyonic sytem into the future. We conclude, that both future evolutions are allowed by observations, Big Brake becoming increasingly likely with the increase of the positive model parameter $k$.
 
\end{abstract}

\maketitle

\section{Introduction}

With the discovery of cosmic acceleration \cite{cosm} the quest for modeling
dark energy \cite{dark} has started. Besides the most simple cosmological
constant, other models based on various perfect fluids with negative
pressure, like Chaplygin gas \cite{Chaplygin}, minimally and non-minimally
coupled scalar fields and fields having non-standard kinetic terms \cite%
{kinetic,tachyons} were advanced. The latter ones include as a subclass the
models based on different forms of the Born-Infeld-type action, which is
often associated with the tachyons arising in the context of string theory 
\cite{string}. Due to the non-linearity of the dependence of the tachyon
Lagrangians on the kinetic term of the tachyon field, the dynamics of the
corresponding cosmological models appears to be very rich.

The tachyon model studied in paper \cite{we-tach} contains a 2-fluid
analogue scalar field $T$, the dynamics of which is given by a simple
potential, depending on two parameters, $\Lambda $ and $k$. The model is
homogeneous and isotropic. A phase space diagram in the tachyonic field and
its derivative $s\equiv \dot{T}$ shows 5 type of distinct cosmological
evolutions possibly occurring for the model, some of them containing regimes
where $s$ is superluminal. All evolutions originate from one of the Big
Bangs of the model, but they either end in a de Sitter infinite exponential
expansion, as the $\Lambda $CDM model does, or in a future singularity
characterized by a regular scale factor $a$, vanishing Hubble parameter $H$
and energy density $\varepsilon $, but infinite $s$ and pressure $p$. Most
notably, the second time derivative of the scale factor goes to $-\infty $,
the reason why we call this singularity a Big Brake.

A kinematical analysis \cite{Barrow} predicted the existence of such
singularities, named sudden future singularities. From a combined
kinematical and observational reasoning alone, sudden future singularities
could occur as early as in ten million years \cite{Dabrowski}, however no
underlying dynamics is known to support this.

Classically the Big Brake singularity is stable.
This can be seen by a series expansion of the scale factor in the
vicinity of the singularity and checking the stability conditions advanced in Ref. \cite{Barrow-priv}. 
Its quantum study indicated singularity avoidance \cite{KKS}.

Recently \cite{tachyon-prd} the compatibility of the model with type Ia
supernovae observation has been investigated. After we present some basic
features of the model in Section 2, in Section 3 we give more details on
this compatibility check, in terms of the original variables employed in
Ref. \cite{we-tach}. Then in Section 4 we stress the crucial difference
between negative and positive values of the model parameter $k$. While for
the former all evolutions end in the de Sitter attractor, for positive $k$
the 1$\sigma $ contour compatible with type Ia supernovae contains both
states which evolve into de Sitter or into a Big Brake. In this dynamical
model the Big Brake can occur no earlier than $10^{8}$ years.

The Big Brake singularity belongs to the class of soft cosmological
singularities, which also includes other representants \cite{soft}. Other
types of singularities arising in the study of various dark energy models
include the Big Rip singularity \cite{Rip}, present in some models with
phantom dark energy \cite{phantom}. The possibility of existence of a phase
of contraction of the universe, ending up in the standard Big Crunch
cosmological singularity was also considered \cite{Crunch}.

\textit{Unit convention: }the Newtonian constant is normalized as $8\pi
G/3=1 $ and we take $c=1$.

\section{The tachyonic model}

We consider the flat Friedmann universe $ds^{2}=dt^{2}-a^{2}(t)dl^{2},$where $dl$ is the spatial distance and $a$ the scale factor, containing a tachyon field $T$ evolving according to the Lagrangian 
\begin{equation}
L=-V(T)\sqrt{1-g^{\mu\nu}T_{,\mu}T_{,\nu}}.  \label{Lagrange}
\end{equation}%
The energy density and pressure of the tachyon field for the Friedmann background are: 
\begin{equation}
\varepsilon =\frac{V(T)}{\sqrt{1-\dot{T}^{2}}}~,\qquad p=-V(T)\sqrt{1-\dot{T}%
^{2}}.
\end{equation}%
We shall consider the model with the tachyonic potential \cite{we-tach}: 
\begin{equation}
V(T)=\frac{\Lambda }{\sin ^{2}\left( \frac{3}{2}\sqrt{\Lambda (1+k)}T\right) 
}\sqrt{1-(1+k)\cos ^{2}\left( \frac{3}{2}\sqrt{\Lambda (1+k)}T\right) },
\label{poten}
\end{equation}%
where $\Lambda $ is a positive constant and $-1<k<1$. The dynamics of the
tachyonic field is encompassed in the system: 
\begin{equation}
\dot{T}=s,  \label{system1}
\end{equation}%
\begin{equation}
\dot{s}=-3\sqrt{V}(1-s^{2})^{3/4}s-(1-s^{2})\frac{V_{,T}}{V},
\label{system2}
\end{equation}%
while gravitational dynamics is given by the Friedmann equation 
\begin{equation}
H^{2}=\varepsilon ,  \label{Friedmann1}
\end{equation}%
where the Hubble variable $H$ is defined as $H\equiv \dot{a}/a$.

For a negative parameter $k$, the evolution of the system (\ref{system1})-(%
\ref{system2}) is always characterized by $-1\leq s\leq 1. $ The evolutions
start from a Big Bang and the system has an attractive node at%
\begin{equation}
T_{0}=\frac{\pi }{3\sqrt{\Lambda (1+k)}},\qquad \ s_{0}=0,  \label{crit}
\end{equation}%
which corresponds to a de Sitter expansion with Hubble parameter $H_{0}=%
\sqrt{\Lambda }$. (For more details see Ref. \cite{tachyon-prd}.)

\begin{figure}[tbp]
\includegraphics[height=6cm, angle=360]{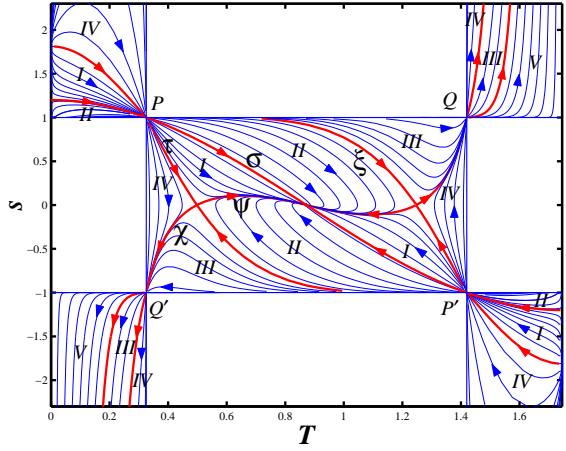}
\caption{Phase portrait evolution for $k>0$ ($k=0.44$).}
\label{fig1}
\end{figure}

The case $k>0$ is much more richer (see Fig. \ref{fig1}). The dynamical
system (\ref{system1})-(\ref{system2}) has three fixed points: the node (\ref%
{crit}) and the two saddle points with coordinates 
\begin{equation}
T_{1}=\frac{2}{3\sqrt{(1+k)\Lambda }}\mathrm{arccos}\sqrt{\frac{1-k}{1+k}}%
,\qquad s_{1}=0,  \label{T1}
\end{equation}%
and, respectively, 
\begin{equation}
T_{2}=\frac{2}{3\sqrt{(1+k)\Lambda }}\left( \pi -\mathrm{arccos}\sqrt{\frac{%
1-k}{1+k}}\right) ,\qquad s_{2}=0,  \label{T2}
\end{equation}%
which give rise to an unstable de Sitter regime with Hubble parameter $H_{1}=%
\sqrt{(1+k)\Lambda /2\sqrt{k}}>H_{0} $.

The most striking feature of the model under consideration with $k>0$
consists in the fact that now the cosmological trajectories are not confined
to the rectangle given by $-1\leq s\leq 1 $ and $T_{3}\leq T\leq T_{4}$,
where 
\begin{equation}
T_{3}=\frac{2}{3\sqrt{(1+k)\Lambda }}\mathrm{arccos}\frac{1}{\sqrt{1+k}},
\label{T3}
\end{equation}%
\begin{equation}
T_{4}=\frac{2}{3\sqrt{(1+k)\Lambda }}\left( \pi -\mathrm{arccos}\frac{1}{%
\sqrt{1+k}}\right)  \label{T4}
\end{equation}%
are the limits of the domain for which the potential $\ V$ is well-defined.
Indeed, the curvature scalar%
\begin{equation}
R=\frac{3V\left( T\right) \left( 4-3s^{2}\right) }{\sqrt{1-s^{2}}}~
\end{equation}%
indicates curvature singularities at $s=\pm 1$ except when $V\left( T\right)
=0$ at the same time. This happens at the points $P,Q,Q^{\prime }$ and $%
P^{\prime }$, where as the analysis of \cite{we-tach} shows, there is no
singularity and the trajectories can be continuated. In doing so, the
potential should be redefined by multiplying with $i$, so that it becomes
real, $W(T)=iV\left( T\right) $. Then in the energy density and pressure $%
\sqrt{1-s^{2}}$ will absorb this $i$, so that in the superluminal regimes we
have 
\begin{equation}
\varepsilon =\frac{W(T)}{\sqrt{s^{2}-1}} , \qquad p=W(T)\sqrt{s^{2}-1},
\label{en-new}
\end{equation}%
both positive.

In what follows, we briefly display all possible classes of cosmological
evolutions existing in the tachyonic model with $k>0$. First of all, note
that in the phase space the reflections with respect to the node point $%
T=T_{0},s=0$ leave the cosmological evolutions invariant. Thus, it makes
sense to study only half of the possible initial conditions in the
rectangle. This rectangle in the phase space $(T,s)$ should be complemented
by four infinite stripes (see Fig.1). The left upper stripe (the right lower
stripe) corresponds to the initial stages of the cosmological evolution,
while the right upper stripe (and the left lower stripe) corresponds to the
final stages. There are five classes of qualitatively different cosmological
trajectories. In characterizing them, we shall consider only half of the
possible initial conditions taking into account the reflection symmetry
mentioned above. 

The trajectories of class IV begin in the left upper stripe in the point
with coordinates $T=0,s=\sqrt{\frac{1+k}{k}}$, which corresponds to the
singularity of the standard Big Bang type. These trajectories climb to some
maximal value of $s$, then turn down and cross the point $P$, entering the
rectangle. Then they leave the rectangle through the point $Q^{\prime }$
entering the left lower stripe. Here, after a finite period of time the
universe encounters a special type of cosmological singularity, which we
call Big Brake. At this singularity, the tachyon field has some finite
value, its velocity $s$ tends to $-\infty $, the cosmological radius has a
finite value, its first time derivative is equal to zero, while its second
time derivative tends to $-\infty $. The trajectories of class I also begin
in the point with coordinates $T=0,s=\sqrt{\frac{1+k}{k}}$, however, after
entering the rectangle they end their evolution in the de Sitter node. They
are separated from the trajectories of class IV by the separatrix $\tau $,
which inside the rectangle connects the corner $P$ with the left saddle
point. The trajectories of class II are separated from those of class I by
the curve $\sigma $, which begins in the point with coordinates $T=0,s=\sqrt{%
\frac{1+k}{k}}$, passes through the corner $P$ and ends in the de Sitter
node. These trajectories begin at the singularity $s=1,T=T_{in}$, where $%
0<T_{in}<T_{\ast },\ T_{\ast }>T_{3}$ and end in the de Sitter node. The
separatrix $\xi $, ending in the right saddle, separates the trajectories of
class II from those of class III. The latter, beginning at $s=1$ and $%
T=T_{in}$, where $T_{\ast }<T_{in}<T_{4}$, after crossing the corner $Q$
encounter their Big Brake singularity in the upper right infinite stripe.
These trajectories are separated from those of class IV by the curve $\chi $%
, which passes through the right saddle point and the corner $Q$. Finally,
the trajectories of the last class V begin at $s=1,T=T_{in}>T_{4}$ and end
in the Big Brake singularity. The time dependence of the Hubble parameter
for these five classes is represented in Fig. 2. 
\begin{figure}[tbp]
\includegraphics[height=6cm, angle=360]{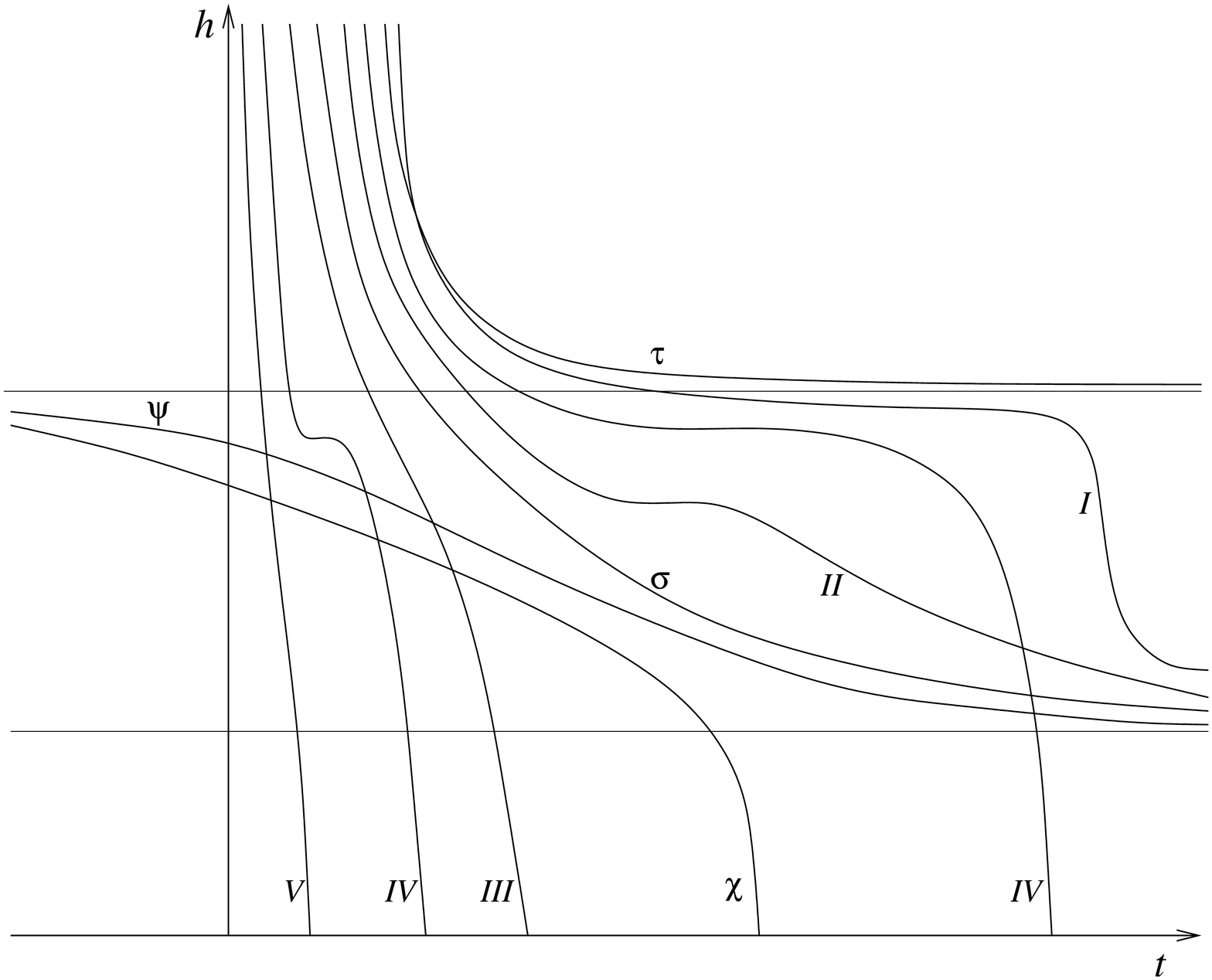}
\caption{Time evolution of the Hubble parameter $H(t)$.}
\label{Fig5}
\end{figure}

We conclude this section by giving some additional formulae characterizing
the different types of singularities present in the cosmological model under
consideration. In the vicinity of the singularity which takes place at the
horizontal sides of the rectangle (say, at $s = 1$), we have the following
dependence of the function $s$ on $T$ \cite{we-tach}: 
\begin{equation}
s = 1 - C(T_{in})(T-T_{in})^4,  \label{singul}
\end{equation}
where 
\begin{equation}
C(T_{in}) = \frac{81}{32}\frac{\Lambda^2\left(1-(1+k)\cos^2\frac{3\sqrt{%
\Lambda(1+k)}T_{in}}{2}\right)}{\sin^4\frac{3\sqrt{\Lambda(1+k)}T_{in}}{2}}.
\label{C-value}
\end{equation}
Hence the energy density is 
\begin{equation}
\varepsilon = \frac{V(T_{in})}{\sqrt{2C(T_{in}}(T-T_{in})^2} = \frac{4}{%
9(T-T_{in})^2}.  \label{en-sing}
\end{equation}
In the vicinity of the singularity $T-T_{in} = t$ and 
\begin{equation}
\varepsilon = \frac{4}{9t^2}  \label{en-sing1}
\end{equation}
while the Hubble variable is 
\begin{equation}
H = \frac{2}{3t},  \label{dust}
\end{equation}
just like in the dust-filled universe born in the vicinity of the Big Bang
singularity.

For the universe born in the point $s = \sqrt{\frac{1+k}{k}}, T = 0$ the
potential $W$ behaves as 
\begin{equation}
W(T) = \frac{4\sqrt{k}}{9(1+k)T^2},  \label{pot-sing}
\end{equation}
where $T = \sqrt{\frac{1+k}{k}}t$. The energy density behaves as 
\begin{equation}
\varepsilon = \frac{4k^2}{9(1+k)^2t^2},  \label{en-sing2}
\end{equation}
while the Hubble variable is 
\begin{equation}
H = \frac{2k}{3(1+k)t}.  \label{Hubble-sing}
\end{equation}
Thus, one can note that the universe has at this point a Big Bang
singularity and behaves in such a way as if it were filled with a perfect
barotropic fluid with equation of state parameter $w = \frac{1}{k}$.

We can also describe the behavior of the universe in the vicinity of the
final Big Brake singularity following the logic of paper \cite{we-tach}.
Consider the universe which is approaching the Big Brake in the lower left
stripe at some value of the tachyon field $T_{BB}$. Correspondingly, the
variable $s$ approaches $-\infty$. Analyzing Eq. (\ref{system2}) in this
limit we have 
\begin{equation}
|s| = \left(\frac{4}{81 W(T_{BB})}\right)^{1/3}(t_{BB}-t)^{-2/3},
\label{s-BB}
\end{equation}
where $t_{BB}$ means the moment of Big Brake. Now, using the formula (\ref%
{Friedmann1}) and the energy density from (\ref{en-new}), one easily finds 
\begin{equation}
H = \left(\frac{9W^2(T_{BB})}{2}\right)^{1/3} (t_{BB}-t)^{1/3}.  \label{H-BB}
\end{equation}
Thus, we see that when $t \rightarrow t_{BB}$, the Hubble variable $H$
vanishes while its time derivative diverges, tending to $-\infty$. It is
important to emphasize that the value $T_{BB}$ is rigorously positive $%
T_{BB} > 0$ \cite{we-tach}.

\section{Confrontation with type Ia supernovae}

\begin{figure}[ht]
\includegraphics[height=8cm, angle=270]{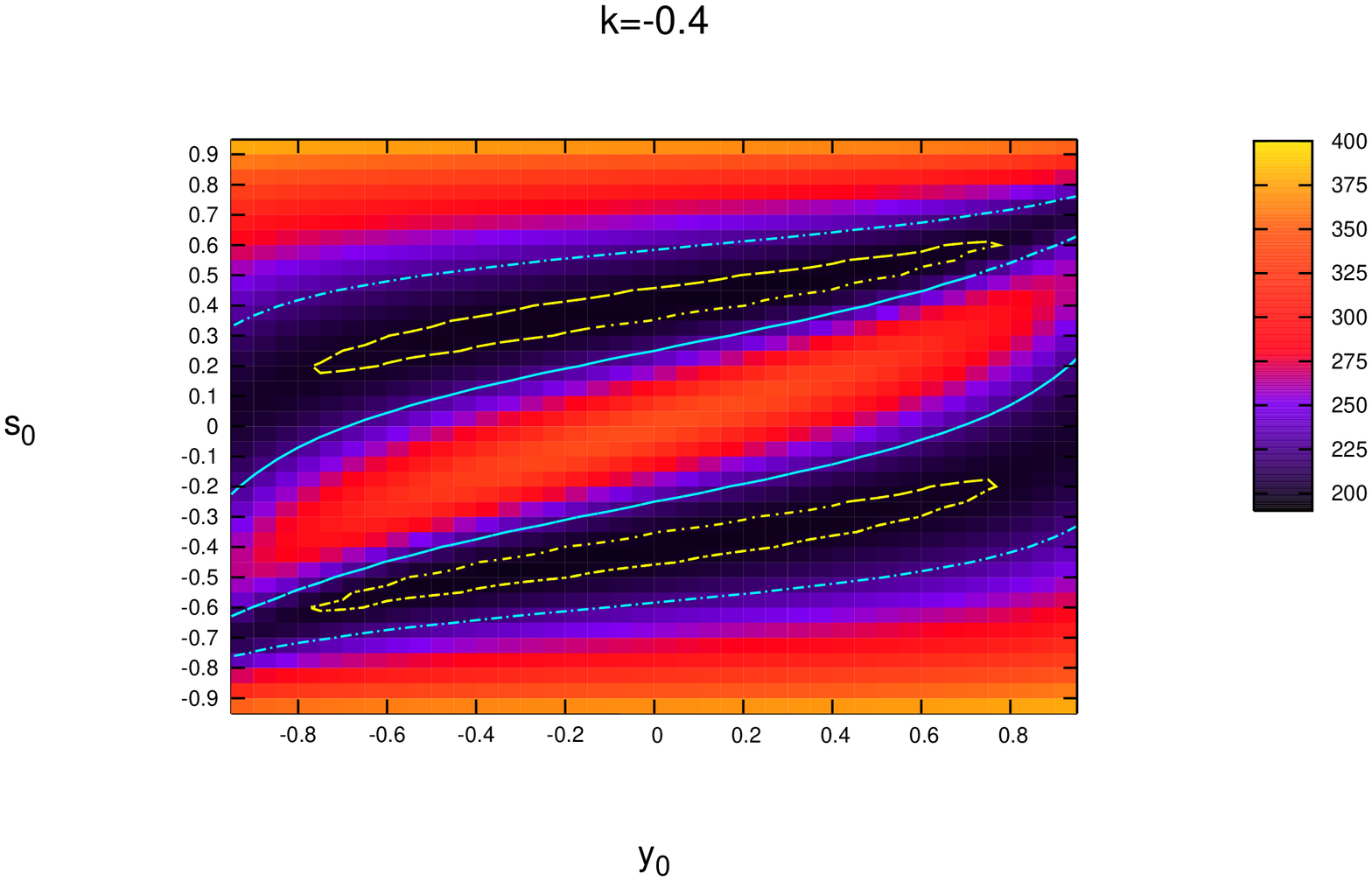} \hskip 1cm %
\includegraphics[height=8cm, angle=270]{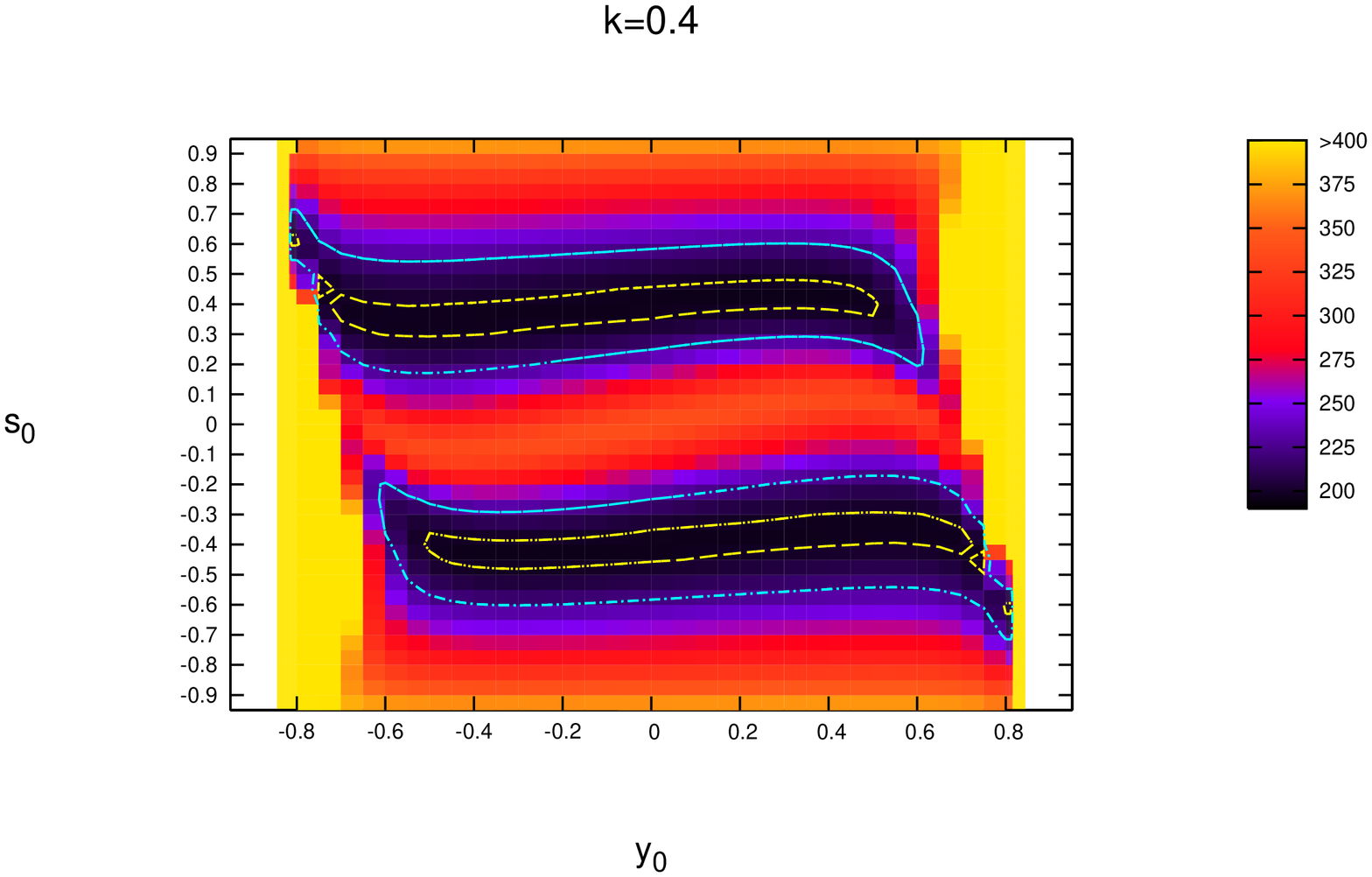}
\caption{ The fit of the luminosity distance vs. redshift for $k=-0.4$
(left) and $0.4$ (right). The white areas represent forbidden parameter
regions where the tachyonic field would be superluminal today. The contours
refer to the $68.3\%$ (1$\protect\sigma$) and $95.4\%$ (2$\protect\sigma$)
confidence levels. For increasing values of $\left\vert k\right\vert <1$%
\thinspace\ the well-fitting regions are increasingly smaller \protect\cite%
{tachyon-prd}. The colour code for $\protect\chi^2$ is indicated on the
vertical stripes. The model is symmetric under the simultaneous change of
signs $y_0\to -y_0$ and $s_0\to -s_0$, thus there is a double coverage of
the parameter space.}
\label{fig3}
\end{figure}

Following Ref. \cite{DicusRepko}, in Ref. \cite{tachyon-prd} we have
presented in detail how to perform a $\chi ^{2}$-test for comparing the
prediction of the model with the available type Ia supernovae taken from
Ref. \cite{SN2007}. In order to do this, we introduce more suitable
dimensionless variables 
\begin{equation}
\hat{H}=\frac{H}{H_{0}},\,\hat{V}=\frac{V}{H_{0}^{2}},\,\Omega _{\Lambda }=%
\frac{\Lambda }{H_{0}^{2}},\,\hat{T}=H_{0}T,  \label{new-var}
\end{equation}%
where $H_{0}$ is the present value of the Hubble parameter. In general, for
any variable $f\left( z\right) $ we will denote by $f_{0}=f(z=0)$. As a
follow-up, we also introduce a new tachyonic variable 
\begin{equation}
y=\cos \left( \frac{3}{2}\sqrt{\Omega _{\Lambda }(1+k)}\hat{T}\right) ,
\end{equation}%
and switch from the time derivative to the derivative with respect to the
redshift $z$ by 
\begin{equation}
\frac{d}{dt}=-H(1+z)\frac{d}{dz}.  \label{der-z}
\end{equation}%
Then we rewrite the equations (\ref{Friedmann1}), (\ref{system1}), (\ref%
{system2}) in terms of the new variables $\hat{H},~s,~y$ and perform the $%
\chi ^{2}$-test. For this we employ%
\begin{equation}
\frac{d}{dz}\left( \frac{\hat{d}_{L}}{1+z}\right)=\frac{1}{\hat{H}}\ ,
\label{dLz}
\end{equation}%
where $\hat{d}_{L}=H_{0}d_{L}$ and $d_{L}$ is the luminosity distance for a
flat Friedmann universe: 
\begin{equation}
d_{L}\left( z\right) =\left( 1+z\right) \int_{0}^{z}\frac{dz^{\ast }}{%
H\left( z^{\ast }\right) }\ 
\end{equation}%
The results are represented on the figure panel \ref{fig3}


\section{Future evolution}

\begin{figure}[ht]
\includegraphics[height=8cm, angle=270]{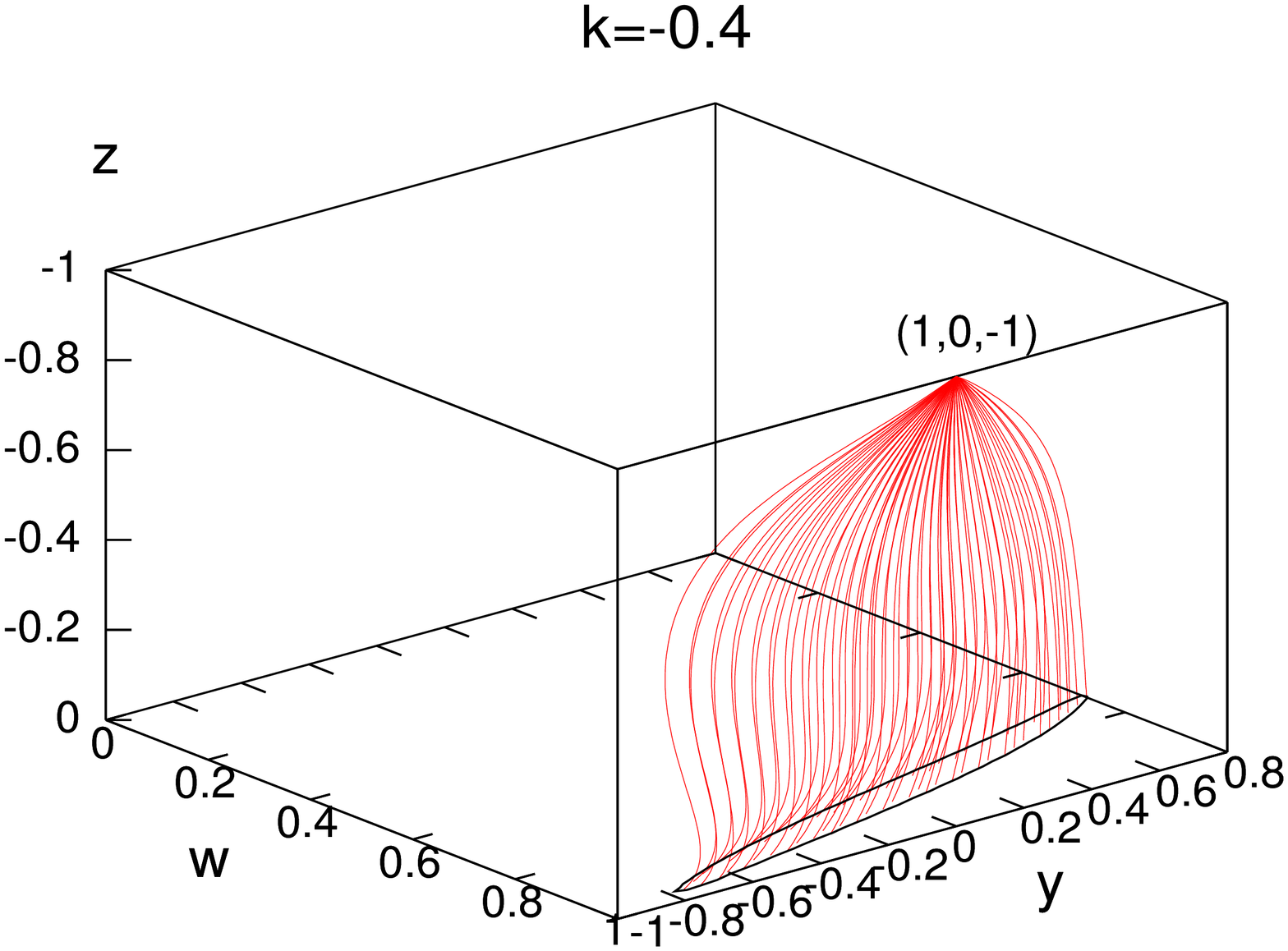} %
\hskip 1cm %
\includegraphics[height=8cm,angle=270]{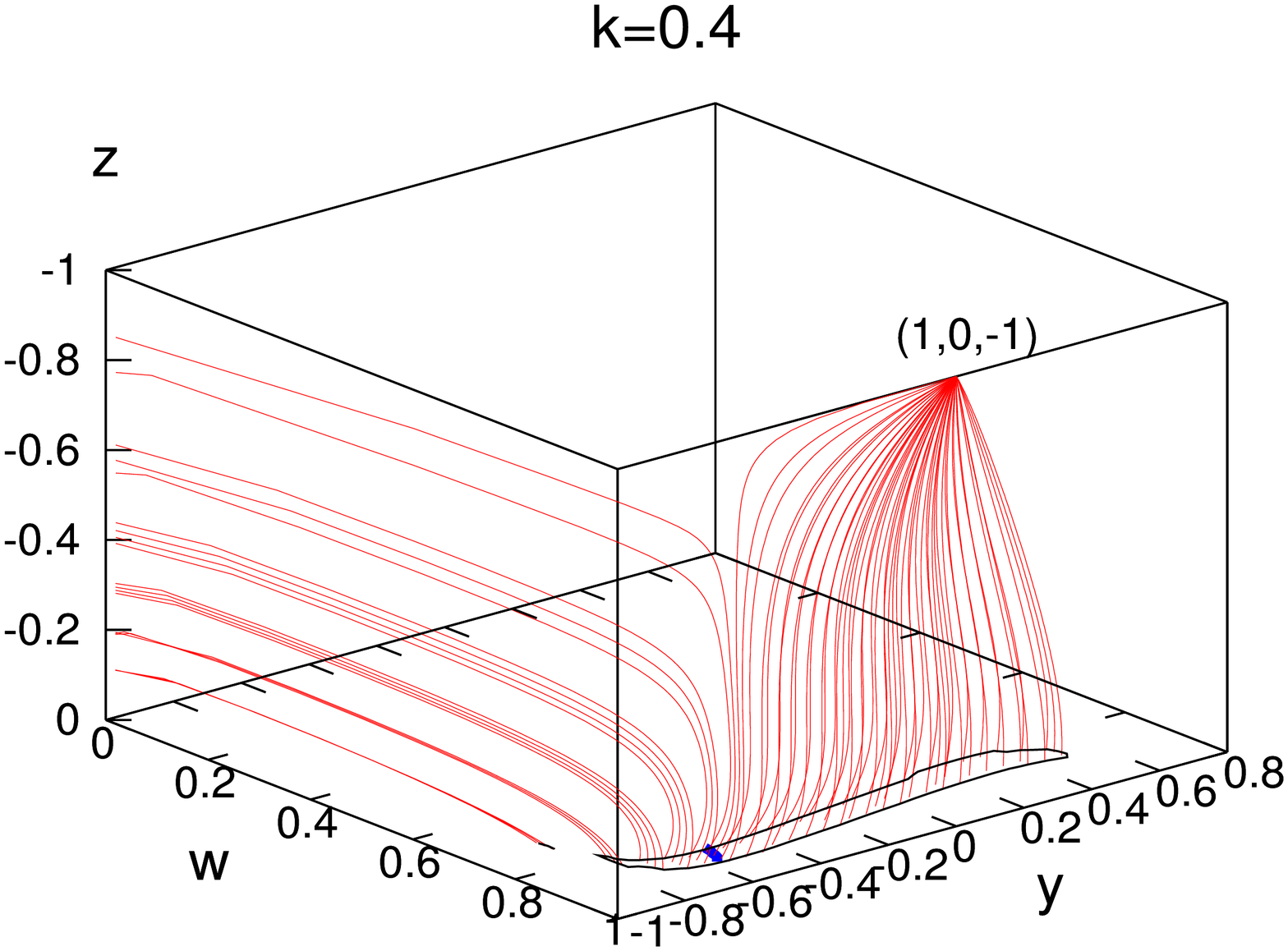}
\caption{The future evolution of those universes, which are in a $68.3\%$
confidence level fit with the supernova data. The 1$\protect\sigma $
contours (black lines in the $z=0$ plane) are from Fig \protect\ref{fig3}
(the parameter plane $\left( y_{0},w_{0}\right) $ is the $z=0$ plane here).
The figures are for $k=-0.4$ (left) and $k=0.4$ (right). The coordinate $w$
is related to the tachyonic speed as $w=1/\left( 1+s^{2}\right) $). In the $%
k=0.4$ case from the 1$\protect\sigma $ parameter range the universe evolves
either into a de Sitter regime or towards the Big Brake singularity. }
\label{fig4}
\end{figure}

In order to avoid the double coverage of the parameter space; also to bring
the Big Brake at $s\rightarrow \pm \infty $ to finite parameter distance, we
introduce the new variable 
\begin{equation}
w=\frac{1}{1+s^{2}}~.  \label{w0}
\end{equation}%
We do this by numerical integration of equations of motion from $z=0$
towards negative values of $z$. We represent the future evolution for $k=\pm
0.4$ on Fig \ref{fig4}. The evolution curves start from the allowed region ($%
w_{0},y_{0}$) in the plane $z=0$. The final de Sitter state is characterized
by the point ($w_{dS}=1,y_{dS}=0,z_{dS}=-1$), the Big Brake final state by
points ($w_{BB}=0,-1<y_{BB}<0,-1<z_{BB}<0$).

Whereas all trajectories with $k=-0.4$ end up eventually into the de Sitter
state, those with $k=0.4$ can either evolve into the de Sitter state or into
the Big Brake state, depending on the particular initial condition ($%
w_{0},y_{0}$). These are generic features holding for negative and positive
values of $k$, respectively. In Ref. \cite{tachyon-prd} we have also found,
that future evolutions towards the Big Brake singularity of the universes
selected by the comparison with supernovae data become more frequent with
increasing (positive) $k$.


For all future evolutions encountering a Big Brake singularity we have
computed the actual time $t_{BB}$ it will take to reach the singularity,
measured from the present moment $z=0$, using the equation $\left(
H_{0}t\right) ^{\prime }=-\hat{H}^{-1}\left( 1+z\right) ^{-1}$). The results
are shown in Table \ref{table1}. The parameter values at which the pressure
turns from negative to positive (at the superluminal crossing) are also
displayed.

\begin{table}[t]
\caption{Properties of the tachyonic universes with $k=0.4$ which (a) are
within 1$\protect\sigma $ confidence level fit with the type Ia supernova
data and (b) evolve into a Big Brake singularity. Columns (1) and (2)
represent a grid of values of the allowed model parameters. Columns (3) and
(4): the redshift $z_{\ast }$ and time $t_{\ast }$ at the future tachyonic
crossing (when $s=1$ and the pressure becomes positive). Columns (5) and
(6): the redshift $z_{BB}$ and time $t_{BB}$ necessary to reach the Big
Brake. The former indicates the relative size of the universe when it
encounters the Big Brake. (The values of $t_{\ast }$ and $t_{BB}$ were
computed with the Hubble parameter $H_{0}=73$ km/s/Mpc.)}
\label{table1}
$%
\begin{tabular}{c|c|c|c|c|c}
$y_{0}$ & $w_{0}$ & $z_{\ast }$ & $t_{\ast }\left( 10^{9}yrs\right) $ & $%
z_{BB}$ & $t_{BB}\left( 10^{9}yrs\right) $ \\ \hline
$-0.80$ & $0.710$ & $-0.059$ & $0.8$ & $-0.106$ & $1.6$ \\ 
$-0.80$ & $0.725$ & $-0.059$ & $0.8$ & $-0.105$ & $1.6$ \\ 
$-0.80$ & $0.740$ & $-0.060$ & $0.8$ & $-0.105$ & $1.6$ \\ 
$-0.75$ & $0.815$ & $-0.144$ & $2.1$ & $-0.184$ & $2.9$ \\ 
$-0.75$ & $0.830$ & $-0.147$ & $2.2$ & $-0.187$ & $3.0$ \\ 
$-0.75$ & $0.845$ & $-0.150$ & $2.2$ & $-0.189$ & $3.0$ \\ 
$-0.70$ & $0.845$ & $-0.241$ & $3.8$ & $-0.276$ & $4.6$ \\ 
$-0.70$ & $0.860$ & $-0.248$ & $4.0$ & $-0.282$ & $4.7$ \\ 
$-0.70$ & $0.875$ & $-0.256$ & $4.1$ & $-0.290$ & $4.9$ \\ 
$-0.70$ & $0.890$ & $-0.264$ & $4.2$ & $-0.298$ & $5.0$ \\ 
$-0.65$ & $0.860$ & $-0.358$ & $6.2$ & $-0.387$ & $7.0$ \\ 
$-0.65$ & $0.875$ & $-0.372$ & $6.5$ & $-0.400$ & $7.2$ \\ 
$-0.65$ & $0.890$ & $-0.388$ & $6.8$ & $-0.415$ & $7.6$ \\ 
$-0.65$ & $0.905$ & $-0.406$ & $7.2$ & $-0.432$ & $8.0$ \\ 
$-0.60$ & $0.875$ & $-0.521$ & $10$ & $-0.542$ & $11$ \\ 
$-0.60$ & $0.890$ & $-0.551$ & $11$ & $-0.571$ & $12$ \\ 
$-0.60$ & $0.905$ & $-0.587$ & $12$ & $-0.605$ & $13$ \\ 
$-0.55$ & $0.875$ & $-0.756$ & $19$ & $-0.766$ & $20$ \\ 
$-0.55$ & $0.890$ & $-0.837$ & $25$ & $-0.845$ & $26$%
\end{tabular}%
$%
\end{table}
In Ref. \cite{tachyon-prd} we have also shown that the Big Brake final fate
becomes increasingly likely with the increase of the positive model
parameter $k$.

\begin{theacknowledgments}
L\'{A}G was supported by the Hungarian Scientific Research Fund (OTKA) grant no. 69036, the Pol\'{a}nyi and Sun Programs of the Hungarian National Office for Research and Technology (NKTH) and the Institute for Advanced Study, Collegium Budapest. ZK was supported by the OTKA grant 69036. A.K. was partially supported by RFBR grant No. 08-02-00923 and by the grant LSS-4899.2008.2.
\end{theacknowledgments}

\end{document}